# Persuasive Technology Contributions Toward Enhance Information Security Awareness in an Organization

Hani A. Qudaih[1], Mohammed A. Bawazir[2], Shuaibu Hassan Usman[3], Jamaludin Ibrahim[4]

*Faculty of Information and Communication Technology,*
*International Islamic University Malaysia,*
*P.O. Box 10, 50728 Kuala Lumpur*

**ABSTRACT:** *Persuasion is part and parcel of human interaction. The human persuaders in society have been always exit, masters of rhetoric skilled of changing our minds, or at least our behaviors. Leaders, mothers, salesmen, and teachers are clear examples of persuaders. Persuaders often turn to technology and digital media to amplify their persuasive ends. Besides, our lives and how we lead them influenced by technologies and digital media, but for the most part, their effects on our attitudes and behaviors have been incidental, even accidental. Although, nowadays, the use of computers to sell products and services considered as the most frequent application of persuasive technology. In this short paper, based on an extensive review of literatures, we aim to give a brief introduction to persuasive technology, and how it can play a role and contribute to enhance and deliver the best practice of IT. Some challenges of persuasive technology have been discussed. At the end, some recommendations and steps should be taken place to empower IT professional practices have been listed.*

## 1. INTRODUCTION

The technology has changed dramatically in the last twenty years particularly the field of information and communication technologies (ICTs). In today's digital age where we live and work, people and businesses find ICTs very important for carrying out their routine activities and tasks. At the same time, however, they may likely to suffer from security breaches of the ICTs, such as stealing information, blocking or denying their users/clients from services, system failure, or data corruption, breaking and hacking into their computers or networks, and so on. These security breaches have put pressure to organizations to adopt and implement adequate security procedures, policy, and techniques in addition with the diverse management skills to assert controls in both operational and technical aspects to secure ICTs resources.

Some previous studies indicated that securing ICTs resources require strong security programs involving education, training, and awareness apart from having strong IT security procedures, policy, and techniques in the organization [1, 2] .Wolmarans (2003) asserted that it is not enough to have good policies and procedures without efficiently communicating, teaching, and reinforcing them to the staff of the organization. Simply because the organization need to stop assuming that employees can read and understand the published policies and procedures of securing ICTs gadgets [2].

Therefore, this study is aimed to discuss the information security awareness in relation to applying the persuasive technologies. The reason is most organizations are facing challenges that ranges from complex and fast growing business requirements (e.g. sales data, profitability data, and client details), provide access to organization's infrastructures (e.g. hardware and software to remote users) keep ahead from competitors, to meeting a number of legal & regulatory compliances (e.g. yearly report on financial status). All these challenges require information security awareness to the employees of organizations in order to keep them secret and protected from authorized access. The essence of information security awareness programs is to put the stage for training through changing attitudes of the staff in organization to realize the significance of security and the adverse effects of its failure [2] and remind the employees the best procedures to be followed in securing vital information [3]. In addition, organizations that did not handle information security training end up in putting the enterprise at great risk because security of information resources is more to a human issue as it is to a technology issue [4]. Hence, persuasive technology will play a vital role in the increase of information security awareness in organizations. As the organizations are constantly challenged with situations in which they need to persuade people (customers, employees, regulators, etc.) to increase the awareness of information security [5], persuasive technology is poised to help the organizations to address the situation. It is asserted that for an awareness program to be successful and effective, organizations need to target the user behaviors [1, 2], people's attitude or mindset towards change, effective change management plan [2], Risk Analysis, Policies, Procedures, Standards, and Business Continuity Planning [6]. Therefore, this study focus on discussing the ways to which persuasive technology can be used to enhance this awareness program in respect of information security. Literature review will be the basis for this study to explain and discuss the persuasive technology to enhance this information security awareness program in organizations.

## 2. LITERATURE REVIEW
### 2.1. Overview of Persuasive Technology

Twenty years ago, few of persuasive technologies were exist in our lives. The web wasn't ubiquitous, and software wasn't designed to change peoples' behaviors; its main concentration was more on crunching data and boost up productivity [7]. Trying to convince persuasion is an important part and parcel of human interaction. Persuasive efforts abound in a continuous attempt to influence our values, attitudes, or behaviors, and convincing us to spend more money on one merchandise rather than another.

Nowadays, however, persuasive technologies are ubiquitous, technology becomes an especially powerful tool when it allows the persuasive techniques to be interactive rather than one-way. Digital media technology, such as digital video, augmented reality, digital audio, or digital art, has played a trivial role in facilitating the delivery of persuasive messages to change what we think and do [8]. As a result, our way of thinking and doing things have been affected because of digital products that circled us. Internet services, mobile devices, desktop computers, and video games considered as interactive, computational technologies which the most self-identified persuasive technology research focus [9].

Despite the fact that, persuasive technologies are found in many domains including Human Computer Interaction (HCI), education, advertising, and health care promotion. A huge recent attention has focused on behavior change in information security domains. Information security domains provide the foundation for security practices and principles in all industries. Such domains included Cryptography, Security Management Practices, Telecommunications and Networking Security, Operations Security, Physical Security, Business Continuity, and Disaster Recovery Planning.

### 2.2. Persuasive Technology Usage in Organizations

Changing peoples' beliefs, values, and actions within an organization is incredibly difficult. An internal or external consultant within the organization may play a unique role to persuade people to deliver a high level of knowledge sharing as well as driving successful change across the globe [10]. Support the specific solution development and expertise or support the project management in the organization not the only main roles of consultants, but its bond of that; they are often act and present a key player in the change management activities that support project implementation [11].

Managers within organizations still be likely to think that staff whom are manage, employed, and paid to do business duties should follow the instructions and the rules they made. At a very early age we are conditioned to believe that persuade people within organizations is the way to train, teach, and to motivate people towards changing what they do. To implement these changes it requires that many people have to take-on of new values and beliefs and the performance of new functions, sometimes numbering in the thousands of people and sometimes spread around the globe. Driving this type of deep change in such diverse situations and large numbers is extremely challenging.

### 2.3. Information Security awareness in an Organization

Information considered as the forms the backbone of organizations as well as it's one of the most and the critical business' assets. But the security of that asset is often overlooked, which is why over 80% of security breaches come from within the organization as a result of poor policy, procedures and staff awareness training [12]. Nowadays, one of the best practices within organizations to protect their information from attacks and breaches is to adapting and complying information security standards such as ISO/IEC 27001:2013 standard. The standard provide at least a baseline set of controls which cover the places, people, and process requirements that organizations need in order to provide suppliers, staff, and customers with confidence in its data security.

Moreover, High-risk data is any information that, when lost, can lead to significant contractual or legal liabilities; serious damage to your organization's image and reputation; or legal, financial, or business losses. In a business, personal data protection is become essential and curtail things. A company should guarantee the confidentiality of the data from the individual which it can bring an issue to the development of the business. For instance, getting the wrong person as an expert in accounting and can't manage timely manner to reconcile accounts from the customers it can cause a bad reputation for the business. Moreover, a company should follow the right regulation from the business rule to avoid fine either from the customer as an individual or the government as a provider of the regulation. For example, in Airlines Companies, when passengers want to do some changing in their travels or tickets, we should re-confirmation about their data again to avoid errors in execution and complaints from the passengers. Thus, the personal data from the individual should be considered of the business protection in the market. Using an experienced expert consultant and auditor in the company is one of the best ways to protect the business from some issues such as bankrupt and damage reputation.

## 3. PERSUASIVE TECHNOLOGY STRATEGIES

The real vital factor to success of information security is effective user security behaviors [13] .As present techniques of helping users and organizations to create more information security have proven unsuccessful, we suggest a set of persuasive technology strategies that persuading others and is used to a certain degree in most everyday interaction and dealing with information security. Our objectives through these strategies are to help, influence, encourage, motivate and educate users to increase the awareness of the significance of information security and the consequences and real threats to users and their organization as well as how to behave securely and utilize coping strategies [5]. For each strategy, it has been identified the features of practical security as following:

- **Simplification**: Contains decrease the processes to the minimum number of actions, beside that reducing the complication of the residual tasks. Simplifying the security responsibilities for users will help them to form a precise mental model of successful processes in information security tasks [5].

- **Tunneling:** It is a process or experience providing opportunities which guide users and motivate them along the way [13]. Using a sequence of tasks to make sure that users followed each step of the intervention process. That would be improved the information security awareness.

- **Personalization:** Means personalized information for each users, normally provides a more personal and attractive experience, which can be more convincing than general information. By providing timely personalized guidance relating to the individual's preferences, requirements, or context-of-use, the system can offer particular details about why current behavior of certain user is insecure and how it can be improved to be more safe and secure, which also increase the understanding of how behave securely and improve users' mental model of information security [5].

- **Monitoring:** Monitoring user status or performance can report it directly to the user himself, who possibly will correct their behavior in harmony with security rules. As a result, that offer the opportunity for users to absorb what should be done to start performing more securely [5]

- **Conditioning:** There is a little direct experience with many users about the consequences and potential threats and risks to the system. They need to encourage them to the correct behavior because the workplace environment does not support users. By applying different methods of reinforcement can help form the desired behavior or change current behaviors into more secure habits [5].

## 4. THE USE OF PERSUASIVE TECHNOLOGY TO ENHANCE INFORMATION SECURITY AWARENESS PROGRAM

Information security requires the elements of standards, procedures, policies, and guidelines within the organization. In addition, it requires the change management that can influence a cultural change and attitude of the staff in an organization. Notwithstanding, organizations that have information security standards, procedures, policies, and guidelines are just starting of an effective information security program. Strong security architecture involves a process in place to make certain that the employees are made aware of their rights and responsibilities with regard to securing the information assets of the organization [6]. In this area, the persuasive technology plays important role in creating Information security awareness for the organizations. The persuasive technology is poised to help the organizations to address the information security awareness program. It was asserted that for an awareness program to be successful and effective, organizations need to target the user behaviors [1, 2], people's attitude or mindset towards change, effective change management plan [2], Risk Analysis, Policies and Procedures, and Business Continuity Planning [6]. Hence, this section discusses on the ways to which the persuasive technology help the organizations to achieve successful and effective information security awareness program.

### 4.1. Behavioral Change

Security of information asset is critical for organizations to protect it from any threats or compromise. To be successful in information security, an awareness program needs to target the user behaviors [1]. Moreover, organizations need to create awareness on the factors that cause threats and compromises to the information and digital assets as well as how to prevent them to their employees. The employees within organizations must be made aware of the consequences of information and digital assets comprise as well as the importance of keeping the information and digital assets secure away from threats. The reason is that users are weakest link when it comes to information security [2]. It may not be deliberate for the users to be in that position but it can be from the failure of the organizations to implement and create awareness, policy, technology, and training appropriately to prevent them from intentionally or accidentally permitting threats, damage or loss of information asset [2]. Providing the environment for employees to be aware of their responsibility toward information security and training them on the correct practices of it help employees to change their behavior.

Organizations can use the persuasive technology to create awareness and training to their employees toward information security, which can help employees to change their behavior. Persuasive technology is fundamentally about learning to automate behavior change [14] to which can be effectively encoded in creating experiences that change behaviors in information security awareness in an organization. The tools for creating persuasive products have become very easy to be used within organization [14]. As a result, organizations can now design experiences and innovations in social networks, online video, and presentation that influence people's behaviors by means of technology channels. In the part of information security awareness within organization, the implications will be normative influence and social facilitation among the employees. The reason is that people respond socially to computer products has significant implications for persuasion [15].

### 4.2. Change of Attitude

The primary goal for the organizations in creating an information security awareness program is to change the employees' attitude towards information security. Because creating security awareness, give way to a change in attitude that leads to a change in behavior of the employees [2]. Proper information security awareness programs put the stage for educating and training the employees that change their attitudes to become conscious of the importance of security and the adverse consequences of its failure and remind users of the procedures to be followed [2].

Persuasive technology also can use to change attitude by conveying social presence and persuasion. For example, dialogue boxes can be used to persuade users to update software, to stop visiting malicious web sites, to renew password etc. With all these, users may infer that the computing product is animate in some way to which can lead to their attitudes and their behavioral change. Moreover, persuasive technology stimulates and motivates those being trained to care about security and to remind them of important security practices such logging off a computer system or locking doors (Whitman, 2013).

Persuasive technology is an interactive technology that attempts to change attitudes in some way because it enables organizations to create, distribute, or adopt the technology. Actually, the persuasive technology function in broad aspects: as tools, as media, and as social actors [16] . Table 1 depicts the three functions of the Persuasive technology. These three functions of persuasive technology allow people view or respond to computing technologies. As a tool, the persuasive technology (the computer application or system) provides humans with new ability or power, allowing people to do things they could not do before, or to do things more easily [17].

On the other hand, persuasive technology functions as media that can convey either symbolic content (e.g., text, data graphs, and icons) or sensory content (e.g., real-time video, simulations, and virtual worlds. In the aspect of social actor, the persuasive technology enables animation characteristics (physical features, emotions, voice communication), play animation roles (coach, pet, assistant, opponent), or follow social rules or dynamics (greetings, apologies, turn taking) [18]. Therefore, persuasive technology can be used as tool, media, and social actor to influence the attitude of employees in information security awareness in an organization.

| SN | Function | Essence | Persuasive affordances |
|---|---|---|---|
| 1 | Persuasive technology as a tool | Increases capabilities | • Reduces barriers (time, effort, cost)<br>• Increases self-efficacy<br>• provides information for better decision making<br>• changes mental models |

| 2 | Persuasive technology as a medium | Provides experiences | • Provides first-hand learning, insight, visualization, resolve<br>• Promotes understanding of cause/effect relationships<br>• Motivates through experience, sensation |
|---|---|---|---|
| 3 | Persuasive technology as a social actor | Creates relationship | • Establishes social norms<br>• Invokes social rules and dynamics<br>• Provides social support or sanction |

Table 1: Three persuasive technology functions and their persuasive affordances. Source: adapted from [16]

### 4.3. Change Management Plan

The success and effective of information security awareness depends on the relevant procedures and measures taken to enforce it on the overall users of information assets in the organization. Despite having adequate security awareness explained to the users on the importance of information assets, action to be taken or countermeasures, threats, and responsibility concept, several problems can still exist. The biggest problem is people's attitude towards change that requires an effective change management plan must be in place [2].

In this context of change management approach to information security awareness, the persuasive technology can enable the organizations to create security awareness in visual, audio, or textual experiences to their staff. In this way, the persuasive technology can is seen as the powerful digital tool that influences human behavior. For instance, it has been revealed that the current software applications, mobile devices, and Web sites can be used to change people's attitudes and behavior. The author believe that the computers in the present days are playing on a variety of roles as persuaders due to their migration from research labs to desktops and now into everyday life. The author laments that the computers have become more persuasive by design, which change people's attitudes and behaviors [19]. The practitioners, technology designers, researchers, and marketers are leveraging the persuasive power of interactive technology to achieve their goals

Change management approach is crucial in creating of information security awareness as it helps in cultural change. The aim of using the change management in information security awareness is to ensure that security awareness initiative objectives are met, as well as followed within the organization [20]. Therefore, the abilities of the persuasive technology to combine data with graphics, animation, audio, video, simulation, or hyperlinked content create the optimum persuasive impact that match people's preferences [21] that is vital for use in Change management approach.

### 4.4. Risk Analysis

Risk analysis is the process involves assessing and identifying the factors that may endanger the success of achieving information security goals within organization. Therefore, organizations should employ the risk assessment to determine the factors that lead to existence of threats or damages to information asset and the try to associated risk level of those threats to their Information security awareness programs. This will enable the organizations to implement appropriate control measures, safeguards, or counter measures to lower the risk to an acceptable level [6].

Persuasive technologies play a vital role on the described risk management approaches [19]. As a result of incorporating persuasive elements into the technologies, appropriate facilitation in creating Information security awareness program can be presented to users at the right time and place. With persuasive technologies, organizations can design ways to disseminate risk associated with the Information security. This way risk-related information is available ubiquitously that eliminate physical delay and time delay. When compare with human persuaders, persuasive technologies allow storage and accessibility to huge amount of data and information. It also presents the data in different format and models, and enable customization styles that are attractive and meaningful the users. Lastly, persuasive technologies allow simulations that provide personal and interactive experiences on virtual environment. Organizations can use Persuasive technologies to simulate various risks including their preventions and penalties for the employees.

### 4.5. Policy and Procedure

The objectives of setting up policies and procedures along with an information security awareness program in an organization are to establish the idea of the requirement expected of all employees in dealing with the enterprise information assets and the consequences of noncompliance [6].

The persuasive technology helps in policy and procedure to change behaviors in information security awareness in an organization. Nowadays, computing technologies generate different reports and information based on the target users [19]. Hence, organizations can the persuasive technology on information security that creates awareness based on the various need of the employees. With the use of persuasive technology in disseminating of policies and procedures, the organizations can achieve a best result in information security awareness program due to distinction between groups of people and presents only information that is relevant to that particular audience. The job category is one way to segment the awareness audience and will provide the presenter with guidelines as to type and duration of the presentation [6]. With persuasive technology, users will be motivated to be aware of their security responsibilities and disseminating correct practices, which lead user's change past behaviors.

### 4.6. Business Continuity Planning

In today's business arena, organizations need to implement an enterprise-wide continuity plan. Business continuity planning (BCP) allows organizations to demonstrate existence of exercising due diligence with processing and storing of information resources and assets. With business continuity planning, organizations can able to show employees, stakeholders, and interested third parties that the continued operation of the enterprise has been addressed and is taken seriously [6].

The persuasive technologies help support business operations and maintain continuity of services for the organizations through protection of data and improved systems availability. Furthermore, persuasive technologies plays an integral part in supporting restore, back up, and archive of critical data, systems and information on demand for the organization.

## 5. CHALLENGES OF INFORMATION SECURITY AWARENESS

Applying an effective information security awareness through persuasive technology can be a challenging task. Even some of good plan programmes can face large obstacles and challenges. Nevertheless, by understanding some of these common challenges, it will support and help to cope them during the design and execution phases of the programme. Some of the challenges that will face information security awareness include:

• Technology moves faster than organizations' staffs realize that. The issue that the information security awareness of staffs is not up to date or not effectively informed of these types of education till it is too late.

• "One-size-fits-all" strategy might be easier to apply and develop, but it will not be effective. Information security awareness programmes sometimes be unsuccessful to segment their users sufficiently and proper messages are not delivered.

• Information security awareness programme takes long period of time to design and build in order to understand users' needs and find the right balance.

• Consistent style, theme and messages delivery are paramount for improving users' awareness, without them would be difficult for the user to involve in the programme.

• Many information security awareness programmes fail to follow up and maintain with users as regular cycle of communication which absolutely effect the feedback from users.

• Lack of resources definitely effect information security awareness as limitation of what it is able to achieve due to budget cut.

• Lack of management support is also one of the most challenging. Management support is an essential aspects of a security awareness programme because managers have the power of effective messages as their roles and responsibilities.

## 6. RECOMMENDATIONS

There are some recommendations that may help the information security awareness of organization's staff to be aware of information security threats and how they respond in such situations which include:

• Information Security Awareness is critical to all level of organizations' staffs. All staffs should be taught the importance of information security, what the rules are that must be followed, and what to do if there is an attack.

• The awareness of Information security is not a onetime effort. It is extremely important to make sure continuing program to increase awareness in staffs on the value of appropriate information security management.

• Using stages and plans to make new tasks and activities of information security in order to be more manageable for all staffs. The whole information security awareness should be an on-going and not a fixed process.

• It is very significant to analyze information security knowledge, interests, and the needs of the target staffs of preparing information security awareness programme.

• The senior management should involve during all phases of the programme and always make sure of their support.

• Provide training and instruction on how the staffs deal with information security in efficient and effective way.

• Notify staffs about current and potential threats to the system of organization and sensitive information contained in them.

• It is vital for a security awareness programme to put emphasis on internal communications. In addition, make sure that crisis or an emergency communication strategy is existence.

- Developing consistency of communications. This will also aid and help create an identity for the programme and build relationship with the staffs.
- It is very important to listen to the staffs and adjust the programme based on their needs.

### 7. CONCLUSION

In this paper, we discussed the contribution of persuasive technology to enhance and the best delivery of information security consultants to aware the people within organizations about the momentousness of information which is considered as the most important asset. This can be happened when the change management influence a cultural change and attitude of the staff in an organization to follow the right standards, procedures, policies, and guidelines. The persuasive technology helps in change management approach, behavior change, motivation, and the attitude change of information security awareness in the organization. There are some persuasive strategies which assist, motivate, influence, and educate users to increase the awareness of the importance of information security and the actual threats and consequences to the users and their organization as well as how to behave securely and utilize coping strategies. The persuasive technology consultants can help to strengthen an organization's management pitches to clients, training and coaching your staff, as well as help you find success. Some recommendations have been highlighted may help Information Security Consultants and organization's staff to be aware of information security threats and how they respond in such situations.